\documentstyle[12pt,aaspp4,epsf]{article}

\def\fun#1#2{\lower3.6pt\vbox{\baselineskip0pt\lineskip.9pt
\ialign{$\mathsurround=0pt#1\hfil##\hfil$\crcr#2\crcr\sim\crcr}}}
\def\lap{\mathrel{\mathpalette\fun <}}
\def\gap{\mathrel{\mathpalette\fun >}}

\def\mass{{\cal M}}

\def\Msolar{{\mass_\odot}}

\def\beq{\begin{equation}}
\def\eeq{\end{equation}}
\def\logm{\log\left(M_*/ M_{\bullet}\right)}
\def\mh{M_{\bullet}}

\lefthead{Triaxial Nuclei}
\righthead{}

\begin{document}

\title{Orbital Structure of Triaxial Black-Hole Nuclei}


\author{M. Y. Poon \& D. Merritt}

\affil{Department of Physics and Astronomy, Rutgers University,
New Brunswick, NJ 08855}

\vspace{.3in}

\centerline {\it Astrophysical Journal, Vol. 549, Number 1, Part 1, Page 192}

\vspace{.2in}



\begin{abstract}
Orbital motion in triaxial nuclei with central point masses, 
representing supermassive black holes, is investigated.
The stellar density is assumed to follow a power law,
$\rho\propto r^{-\gamma}$, with $\gamma=1$ or $\gamma=2$.
At low energies the motion is essentially regular; the major
families of orbits are the tubes and the pyramids. Pyramid orbits
are similar to box orbits but have their major elongation parallel
to the short axis of the figure. A number of regular orbit families
associated with resonances also exist, most prominently the
banana orbits, which are also elongated parallel to the short axis.
At a radius where the 
enclosed stellar mass is a few times the black hole mass, the 
pyramid orbits become stochastic. The energy of transition to this
``zone of chaos'' is computed as a function of $\gamma$ and of the
shape of the stellar figure; it occurs at lower energies in more
elongated potentials. Our results suggest that supermassive
black holes may place tight constraints on departures from axisymmetry
in galactic nuclei, both by limiting the allowed shapes of regular
orbits and by inducing chaos.
\end{abstract}

\clearpage

\section {Introduction}

An important change in our thinking about galaxy dynamics took place
during the last decade, when it was recognized that the central densities of 
early-type galaxies and spheroids are generically very high. 
Evidence for large
central masses came from high-resolution kinematical studies of nuclear
stars and gas, which revealed the presence in roughly a dozen galaxies
of compact dark objects with masses $10^{6.5-9.5}M_{\odot}$, presumably 
supermassive black holes (\cite{for98}).
Observations with HST also demonstrated very high stellar densities
at the centers of early-type galaxies (\cite{cra93}; \cite{fer94}).
Low luminosity ellipticals have density profiles that
increase as unbroken power laws at small radii, $\rho\sim r^{-\gamma}$,
with $\gamma\approx 2$.
But Kormendy pointed out already in 1985 that bright
galaxies also have central brightness profiles that deviate 
systematically from that of an isothermal core.
The significance of this deviation was not recognized for 
ten more years due to an optical illusion associated with 
projection onto the plane of the sky.
A luminosity density that varies as $r^{-\gamma}$ at small 
radii generates a power-law cusp in projection only if $\gamma>1$.
When $\gamma=1$, the central surface brightness is logarithmically
divergent (e.g. \cite{deh93}, Fig. 1), and the surface brightness profile
differs only subtly from that of a galaxy with an isothermal core.
Nonparametric deprojection of the luminosity profiles of bright
galaxies (\cite{mef95}; \cite{geb96})
revealed that they too harbor power-law cusps 
but with indices $\gamma\lap 1$.

While the origin of the power-law cusps is not clear, there are 
hints that they may be associated with black holes.  Steep
cusps, $\gamma\approx 2$, form naturally in stellar systems where
the black holes grow on time scales long compared to crossing times
(\cite{pee72}; \cite{qhs95}).
No universally accepted model has yet been proposed for the origin
of the weak cusps, but we note here one feature that suggests a 
link to black holes. Two galaxies, NGC 3379 and M87, have weak
cusps with well-determined structural parameters and also have
black holes with accurately determined masses. Table 1 gives for
each galaxy the ``break'' radius $r_b$ at which the central power 
law turns over to a steeper outer profile; and also the stellar mass
$M_*$ contained within $r_b$.  In both galaxies, $M_*(r_b)$ is identical
within the uncertainties with $M_{\bullet}$, the mass of the black hole;
this is in spite of a factor of $\sim 30$ difference in $M_{\bullet}$ 
and $\sim 6$ in $r_b$. 
This rough equality, and the exclusive association of weak cusps
with bright galaxies, is consistent with models in which weak cusps
are produced following galaxy mergers by the ejection of stars 
from the nucleus by binary black holes 
(\cite{emo91}; \cite{mak97}; \cite{quh97}).

In a fixed stellar potential,
the gravitational influence of the black hole is limited to stars
with pericenters $r_p\lap r_g$ where $M_*(r_g)= M_{\bullet}$. In an axisymmetric galaxy,
$r_p$ is bounded from below by the orbital angular momentum $L_z$ 
about the symmetry axis and only a small fraction of the stars, most
of which are confined to the 
nucleus, can be strongly affected by the black hole. But the gravitational
influence of a black hole can extend far beyond the nucleus
in a non-axisymmetric galaxy since orbital angular momenta are not
conserved and stars with arbitrarily large energies can pass close to
the center (\cite{gb85}).  In a triaxial potential containing
a central point mass, the phase space divides naturally into three 
regions depending on distance from the center.  In the innermost region,
$r\lap r_g$, the potential is dominated by the black hole and the motion
is essentially regular.
Sridhar \& Touma (1999) and Sambhus \& Sridhar (2000) demonstrated
the regularity of the motion in black hole nuclei with constant
stellar densities, and Merritt \& Valluri (1999) found similar results
 for motion in triaxial models with weak cusps, $\gamma=0.5$.
Here we extend those results to cusps with the steeper power laws
characteristic of most galaxies.
At intermediate radii, the black hole acts
as a scattering center rendering almost all of the center-filling
orbits stochastic.  This ``zone of chaos'' extends outward from a few 
times $r_g$ to a radius where the enclosed stellar mass is roughly $10^2$ 
times the mass of the black hole (\cite{mer99}). 
In the outermost region, the phase 
space is a complex mixture of chaotic and regular trajectories, including
resonant box orbits that remain stable by avoiding the center 
(\cite{caa98}; \cite{pal98}; \cite{vam98}; \cite{waf98}).

The focus of the present study is on the inner two regions and specifically
on the transition from ordered motion near the black hole to chaotic motion
at $r\gap r_g$.  As noted above, the break radius $r_b$ is of order $r_g$
in the two bright elliptical galaxies where both radii can be accurately
measured. The sudden change in the orbital behavior near $r_g$ 
might therefore imply a change in the three-dimensional shapes of galaxies 
near $r_b$.  In fact there is some evidence for changes in ellipticity and 
boxiness in bright elliptical galaxies at $r\approx r_b$ 
(\cite{ryd99}; \cite{qui99}; \cite{bes99}).  
The work presented here is a prelude to full self-consistency studies which 
will place more rigorous constraints on the allowed shapes of triaxial 
black-hole nuclei.

In \S 2 we present our model for the stellar density and calculate the
gravitational potential and forces.
The families of orbits are discussed in \S3, and the transition from
regular to chaotic motion is discussed in \S4.
\S5 sums up.

\section {Mass Model, Potential, Forces}

We model the stellar nucleus as a triaxial spheroid with a power-law 
dependence of density on radius:
\begin{mathletters}
\begin{eqnarray}
\rho_{\star} &=& \rho_{\circ} m^{-\gamma},\\
m^2 &=& \frac{x^2}{a^2}+\frac{y^2}{b^2}+\frac{z^2}{c^2}\label{rho}
\end{eqnarray}
\end{mathletters}
within a bounding ellipsoid $m=m_{max}$.
The isodensity surfaces are concentric ellipsoids with fixed axis 
ratios $a : b : c$;
without loss of generality, we assume $a > b > c$. 
Triaxiality is measured by the quantity $T$ which is defined as
\begin{eqnarray}
T \equiv \frac{a^2-b^2}{a^2-c^2},
\end{eqnarray}
so that prolate galaxies ($b=c$) have $T=1$ and oblate galaxies ($a=b$) 
have $T=0$. 
Since the models are scale-free, we are free to assign a mass of $1$ 
to the central point representing the black hole.
Most of the discussion below will be restricted to models with 
$\gamma=1$ (``weak cusp'') and $\gamma=2$ (``strong cusp''), 
typical of the values of $\gamma$ in bright and faint galaxies respectively.

Our assumption of a power-law density dependence with fixed index $\gamma$
is reasonable for galaxies with strong cusps, in which $\rho\sim r^{-\gamma}$,
$\gamma\approx 2$, even at radii well outside the sphere
of influence of the black hole. In weak-cusp galaxies, the shallow inner
power law ($\gamma\approx 1$) eventually turns over to a steeper dependence
at radii $r\gap r_b$.  However, as argued above, $r_b\approx r_g$,
and we show below that $r_g$ is approximately the radius at which a transition
to chaos occurs. Thus our $\gamma=1$ models are expected to yield an
accurate description of the dynamics of weak-cusp nuclei out to at least the
inner edge of the ``zone of chaos.''

The gravitational potential corresponding to the stars can be 
obtained using Chandrasekhar's theorem (\cite{cha69} , p. 52, 
theorem 12) which says that for a density law that is stratified on similar ellipsoids,
the gravitational potential can be written as 
\begin{eqnarray}
\Phi_{*}({\bf x}) = -\pi abcG \int_{0}^{\infty} \frac{[\psi
(m_{max}^2) - \psi (m^2)]}{\sqrt{(\tau + a^2)(\tau + b^2)(\tau + c^2)}} d\tau
\end{eqnarray}
where
\begin{eqnarray}
\psi(m^2) = \int_{m^2_{max}}^{m^2(\tau)} \rho(m'^2) dm'^2
\end{eqnarray}
and 
\begin{eqnarray}
m^2(\tau)= \frac{x^2}{a^2 + \tau} + \frac{y^2}{b^2 + \tau} + \frac{z^2}{c^2 +
\tau}.
\end{eqnarray}

For the weak-cusp ($\gamma=1$) case, we have
\begin{equation}
\Phi_{*}({\bf x}) = -\pi abcG \int_{0}^{\infty}
\frac{\psi(m^2_{max}) + 2\rho_{\circ}m_{max}
-2 \rho_{\circ} \left( 
\frac{x^2}{a^2+\tau}+\frac{y^2}{b^2+\tau}+\frac{z^2}{c^2+\tau}
\right)^{\frac{1}{2}}
}{\sqrt{(\tau + a^2)(\tau + b^2)(\tau + c^2)}}d\tau
\label{eqnFw} \label{PhiE1}
\end{equation}
while for the strong-cusp ($\gamma=2$) case,
\begin{eqnarray}
\Phi_{*}({\bf x})=
-\pi abc G \int_{0}^{\infty}
\frac{\psi(m^2_{max}) + \rho_{\circ}\ln(m^2_{max})
- \rho_{\circ} \ln{\left( \frac{x^2}{a^2 + \tau} + \frac{y^2}{b^2 + \tau} + 
             \frac{z^2}{c^2 + \tau} \right)}} 
{\sqrt{(\tau+a^2)(\tau+b^2)(\tau+c^2)}}d\tau.
 \label{PhiE2}
\end{eqnarray}
The constant terms, depending on $m_{max}$, will be ignored in what follows since they have no effect on the forces.

For convenience of numerical calculation, 
the potential in the strong-cusp
case may be expressed in terms of a new set of
coordinates $\{r, \mu^{*}, \nu^{*} \}$
which have the following definitions (\cite{dep88}):
\begin{mathletters}
\begin{eqnarray}
r^2 = x^2 + y^2 + z^2, \\
\mu^* = \frac{1}{2}d_1 + \frac{1}{2}\sqrt{d_2}, \\
\nu^* = \frac{1}{2}d_1 - \frac{1}{2}\sqrt{d_2},
\end{eqnarray}
\end{mathletters}
where
\begin{mathletters}
\begin{eqnarray}
r^2 d_1 &=& a^2(y^2 + z^2) + b^2(z^2 + x^2) + c^2(x^2 + y^2), \\
r^4 d_2 &=& [(b^2-c^2)x^2-(c^2-a^2)y^2-(a^2-b^2)z^2]^2
	       +4(a^2-b^2)(a^2-c^2)y^2 z^2.
\end{eqnarray}
\end{mathletters}
In terms of these variables, the stellar potential in the strong-cusp case 
can be written as
\begin{eqnarray}
\Phi_{*}({\bf x})=A\ln r + F_1(\mu^*) + F_1(\nu^*),\label{eqnFs}
\end{eqnarray}
where
\begin{mathletters}
\begin{eqnarray}
A &=& 4\pi G\rho_{\circ}abcR_F(a^2, b^2, c^2), \\
F_1(\tau) &=& \pi G \rho_{\circ}abc\int_{0}^{\infty} 
\frac{\ln(\tau+u)}
{\sqrt{(a^2+u)(b^2+u)(c^2+u)}}du
\end{eqnarray}
\end{mathletters}
and
\begin{eqnarray}
R_F(m, n, q)\equiv \frac{1}{2} \int_{0}^{\infty}
\frac{du}{\sqrt{(u+m)(u+n)(u+q)}}
\end{eqnarray}
is the Carlson elliptic integral (\cite{car88}).

Forces may be obtained in analytical form in Cartesian coordinates
for both the strong and weak cusp cases. 
By taking partial derivatives of (\ref{eqnFw}), the weak-cusp 
force components are found to be
\begin{mathletters}
\begin{eqnarray}
F_{*x}&=&-\frac{2\pi Gabc\rho_{\circ}}{\sqrt{a^2-b^2}\sqrt{a^2-c^2}}
	\ln \left( \frac{f_1}{f_2} \right), \\
F_{*y}&=&-\frac{2\pi Gabc \rho_{\circ}}{\sqrt{a^2-b^2}\sqrt{b^2-c^2}}
\left[ \tan^{-1}\left( \frac{f_3}{f_4} \right)
- \tan^{-1}\left(\frac{f_5}{f_6} \right) \right], \\
F_{*z}&=&-\frac{2\pi Gabc\rho_{\circ}}{\sqrt{a^2-c^2}\sqrt{b^2-c^2}}
	\ln \left( \frac{f_7}{f_8} \right),
\end{eqnarray}
\end{mathletters}
where
\begin{mathletters}
\begin{eqnarray}
f_1 & =& \left( 
x \sqrt{a^2-b^2} \sqrt{a^2-c^2} + abc \sqrt{ \frac{x^2}{a^2}+\frac{y^2}{b^2}
+\frac{z^2}{c^2} }\right)^2-a^4(x^2+y^2+z^2) \\
f_2 & = & a^2 \left( 
(b^2+c^2-2a^2)x^2 + (c^2-a^2)y^2+ (b^2-a^2)z^2 \right) \nonumber \\
   & + & 2a^2 x\sqrt{a^2-b^2} \sqrt{a^2-c^2} \sqrt{x^2+y^2+z^2} \\
f_3 & = & x^2(b^2+c^2) + y^2(a^2+c^2) + z^2(a^2+b^2) -2b^2(x^2+y^2+z^2) \\
f_4 & = & 2y\sqrt{a^2-b^2}\sqrt{b^2-c^2}\sqrt{x^2+y^2+z^2} \\
f_5 & = & (x^2b^2c^2+a^2c^2y^2+z^2a^2b^2)-y^2(b^2-c^2)(a^2-b^2)-b^4(x^2+y^2+z^2) \\
f_6 & = & 2y\sqrt{a^2-b^2}\sqrt{b^2-c^2}\sqrt{c^2b^2x^2+a^2c^2y^2+a^2b^2z^2}\\
f_7 & = & (z\sqrt{a^2-c^2}\sqrt{b^2-c^2}+\sqrt{a^2c^2y^2+b^2c^2x^2+b^2a^2z^2})^2 
-c^4(x^2+y^2+z^2) \\
f_8 & = & c^2\left((b^2-c^2)x^2+(a^2-c^2)y^2+(a^2+b^2-2c^2)z^2 \right) \nonumber \\
   & + & 2c^2z\sqrt{a^2-c^2}\sqrt{b^2-c^2}\sqrt{x^2+y^2+z^2}
\end{eqnarray}
\end{mathletters}

For the strong-cusp forces, we take partial derivatives of (\ref{eqnFs}) 
in Cartesian coordinates (\cite{dep88}) and the force components are given by
\begin{mathletters}
\begin{eqnarray}
F_{*x}&=&-\frac{x}{r}\frac{\partial \Phi_{*}}{\partial r}
	+\frac{2x}{r^2}\frac{(\mu-b^2)(\mu-c^2)}{\mu-\nu}\frac{\partial \Phi_{*}}
          {\partial \mu}
	+ \frac{2x}{r^2}\frac{(\nu-b^2)(\nu-c^2)}{\nu-\mu}\frac{\partial \Phi_{*}}
          {\partial \nu},\\
F_{*y}&=&-\frac{y}{r}\frac{\partial \Phi_{*}}{\partial r}
	+\frac{2y}{r^2}\frac{(\mu-a^2)(\mu-c^2)}{\mu-\nu}\frac{\partial \Phi_{*}}
          {\partial \mu}
	+ \frac{2y}{r^2}\frac{(\nu-a^2)(\nu-c^2)}{\nu-\mu}\frac{\partial \Phi_{*}}
          {\partial \nu},\\
F_{*z}&=&-\frac{z}{r}\frac{\partial \Phi_{*}}{\partial r}
	+\frac{2z}{r^2}\frac{(\mu-a^2)(\mu-b^2)}{\mu-\nu}\frac{\partial \Phi_{*}}
          {\partial \mu}
	+ \frac{2z}{r^2}\frac{(\nu-a^2)(\nu-b^2)}{\nu-\mu}\frac{\partial \Phi_{*}}
          {\partial \nu},
\end{eqnarray}
\end{mathletters}
where
\begin{mathletters}
\begin{eqnarray}
\frac{\partial \Phi_{*}}{\partial r} &=& 4\pi Gabcr^{-1}\rho_{\circ}R_f(a^2,b^2,c^2),\\
\frac{\partial \Phi_{*}}{\partial \mu} &=&
	\frac{2}{3}\pi Gabc\rho_{\circ}R_J(a^2, b^2, c^2, \mu),\\
\frac{\partial \Phi_{*}}{\partial \nu} &=&
	\frac{2}{3}\pi Gabc\rho_{\circ}R_J(a^2, b^2, c^2, \nu),
\end{eqnarray}
\end{mathletters}
and
\begin{equation}
R_J(m, n, q, r) \equiv 
\frac{3}{2}\int_{0}^{\infty}\frac{du}{(u+r)\sqrt{(u+m)(u+n)(u+q)}},
\end{equation}
is the Carlson elliptic integral.

The magnitude of the radial force in the weak cusp case is constant
as a function of distance from the center,
while in the strong cusp case, the force diverges as $r^{-1}$.

The dynamical time $T_D(E)$ is defined below as the period of a 
circular orbit of the same energy in the equivalent spherical potential,
which is defined to have a scale length $a_{ave}=\sqrt[3]{abc}$. 
The energy of a circular orbit in the spherical models is
\begin{mathletters}
\begin{eqnarray}
E_{c}(r)&=&3\pi G\rho_{\circ}ra_{ave} 
		- \frac{GM_{\bullet}}{2r}~~~~~~~~~~~~~~~~~~~~~~~~~(\gamma=1),\label{PhiS1}
\end{eqnarray}
\begin{eqnarray}
E_{c}(r)&=&2\pi G\rho_{\circ}a_{ave}^2\left[2\ln
		\left(\frac{r}{a_{ave}}\right)
		-1 \right]-\frac{GM_{\bullet}}{2r}~~~~(\gamma=2),\label{PhiS2}
\end{eqnarray}
\end{mathletters}
and its period is 
\begin{eqnarray}
T_c(r)= \left[   
\frac{G\rho_{\circ}}{\pi(3-\gamma)}\left(\frac{a_{ave}}{r}\right)^{\gamma}
+\frac{GM_{\bullet}}{4\pi^2r^3}\right]^{-\frac{1}{2}}.
\end{eqnarray}
Given the energy $E$ in the triaxial potential, we set $E_c(r_c)=E$ and 
solve for $r_c$ and $T_c(r_c)$. The latter is equated to $T_D(E)$.

Equations (\ref{PhiS1}) and (\ref{PhiS2}) are based on the same choices
for the zero point of the potential as was made in equations (\ref{PhiE1}) 
and (\ref{PhiE2}).

The largest Liapunov exponent was computed for all orbits in the standard way,
by integrating the equations of motion of an infinitesimal perturbation.
Analytical partial derivatives of the forces may be found for both the weak 
and strong cusp cases. 
In the strong cusp case, the expressions may be simplified using the identity
\begin{eqnarray}
\frac{d R_J(a^2, b^2, c^2, \tau)}{d\tau}
=-\frac{3}{2\tau abc} + \frac{1}{2}\left( \frac{1}{a^2-\tau}
+\frac{1}{b^2-\tau}+\frac{1}{c^2-\tau} \right)R_J(a^2, b^2, c^2, \tau) \nonumber \\
-\frac{1}{2(a^2-\tau)}R_D(b^2, c^2, a^2)
-\frac{1}{2(b^2-\tau)}R_D(a^2, c^2, b^2)
-\frac{1}{2(c^2-\tau)}R_D(a^2, b^2, c^2)
\end{eqnarray}
where
\begin{eqnarray}
R_D(m, n, r) \equiv R_J(m, n, r, r).
\end{eqnarray}

The equations of motion were integrated using the routine ``RADAU'' 
of Hairer \& Wanner (1996).
RADAU is a variable time step, implicit Runge-Kutta scheme which automatically
switches between orders of 5, 9 and 13.
Energy conservation was extremely good; energy was typically 
conserved to a few parts in $10^9$ over 100 dynamical times.

As found also in earlier studies (e.g. \cite{mev96}), a histogram of
Liapunov exponents of orbits at a given energy evolves
toward a characteristic form as the orbital integration time is
increased.  The regular orbits produce a spike at small values of
$\sigma$, $\sigma T_D\lap 10^{-1.1}$, whose location moves toward the 
left roughly as the inverse of the integration time. The chaotic orbits 
produce a broader peak with a well-defined maximum, typically at 
$\sigma T_D\approx 0.2$, and a tail that extends almost to the regular 
orbits. The tail corresponds to weakly chaotic orbits that are trapped 
near regular phase space regions for long periods of time. As the integration 
time is increased, the histogram tends toward two well-separated and narrow
peaks as the chaotic orbits become increasingly indistinguishable.  In what 
follows, the identification of chaotic orbits was
based both these histograms and on the configuration-space pictures 
of orbits.

We henceforth adopt units such that $G=a=\rho_0=1$.

\section {Orbit Families}

Although the stellar distribution in our models is scale-free,
the presence of the black hole imposes a scale.
We expect the orbital population to change systematically with 
energy, i.e. with distance from the black hole.
For each mass model, we defined a grid of energy values as follows.
We first adopted a set of values $M_{\star}/M_{\bullet}$,
the ratio of enclosed stellar mass to black hole mass :
\begin{eqnarray}
\log_{10} \left(  \frac{M_{\star}}{M_{\bullet}} \right ) 
= \{-0.1,~0.0,~0.1,~0.2,\cdots 1.7,~1.8\}. \label{energygrid}
\end{eqnarray}
Each value of $M_{\star}/M_{\bullet}$ defines an 
ellipsoidal surface, $m=m_*$, such that
\begin{eqnarray}
M_{\star}&=&2\pi abc\rho_0m_*^2~~~~~~~~~~(\gamma=1), \\
M_{\star}&=&4\pi abc\rho_0m_*~~~~~~~~~~~(\gamma=2).
\end{eqnarray}
We then defined the energy $E$ corresponding to this shell as
\begin{eqnarray}
\Phi(x_*=am_*,~0,~0).
\end{eqnarray}
Table~\ref{T2} (weak cusp) and Table \ref{T3} (strong cusp) 
give $x_*$, $E$, $\log \left(M_{\star}/M_{\bullet}\right)$, 
and $T_D$ for the mass models with $T=0.5$ and $c/a=0.5$.

We followed the standard practice (\cite{sch93}; \cite{mef96}) 
of defining two sets of initial-condition spaces.  Stationary start
space consists of initial conditions lying on an equipotential surface
with zero velocity.  $X-Z$ start space consists of starting points in 
the $x-z$ plane with $v_x=v_z=0$. In an integrable triaxial potential,
stationary start space generates box orbits while $X-Z$ start space
generates mostly tube orbits. These two start spaces probably contain
most of the orbits in reflection-symmetric triaxial potentials
(\cite{sch93}).

As in any non-integrable potential, orbits may be ranked
in a hierarchy depending on their phase-space dimensionality
(Merritt \& Valluri 1999).
Stochastic orbits fill a five-dimensional region and in configuration
space populate the entire accessible volume within an equipotential
surface.
Regular, non-resonant orbits occupy 3-tori and densely fill some
more restricted volume.
Resonant orbits satisfy a single relation of the form
\begin{eqnarray}
\sum_{i=1}^{3} m_i\omega_i=0
\end{eqnarray}
between the fundamental frequencies $\omega_i$, where the 
$m_i$ are integers, not all of which are zero.
Resonant orbits occupy 2-tori and densely fill sheets in configuration 
space; when stable, they have associated with them families of non-thin
orbits which mimic the shape of the parent thin orbit.
Orbits satisfying two independent resonance relations are reduced in 
dimensionality yet again to closed, or periodic, orbits. 
Periodic orbits are characterized
by a single base frequency in terms of which the frequency of motion
in any coordinate (e.g. $x$, $y$, $z$) can be expressed as an
integer multiple.

We label families of orbits associated with a single resonance by
the integer vector $(m_1,m_2,m_3)$ that defines the resonance.
For orbital families associated with doubly resonant, or periodic, 
orbits we use the notation $\nu_1:\nu_2:\nu_3$,
the ratios of the frequencies in $x$, $y$ and $z$.

\begin{figure}
\plotone{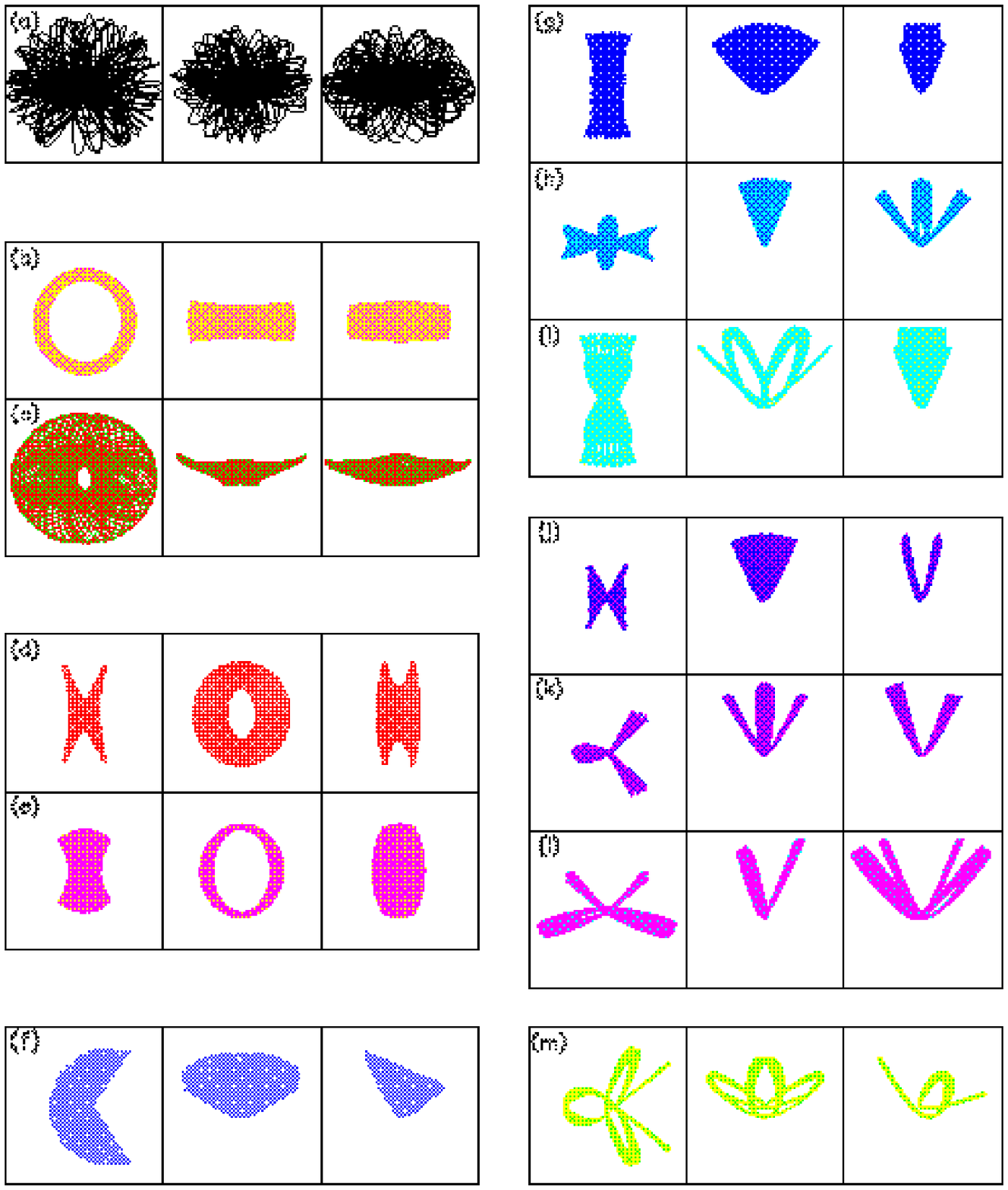} 
\figcaption[fig_orbs.ps]{\label{fig_orbs}}
Major families of orbits in triaxial black-hole nuclei.
Each set of three frames shows, from left to right,
projections onto the ($x,y$), ($y,z$) and ($x,z$) planes.
(a) Stochastic orbit.
(b) Short-axis tube orbit.
(c) Saucer orbit, a resonant short-axis tube.
(d) Inner long-axis tube orbit.
(e) Outer long-axis tube orbit.
(f) ($1,-2,1)$ resonant orbit.
(g) Pyramid orbit.
(h) ($3,0,-4$) resonant pyramid orbit.
(i) ($0,6,-5$) resonant pyramid orbit.
(j) Banana orbit.
(k) $2:3:4$ resonant banana orbit.
(l) $3:4:6$ resonant banana orbit.
(m) $6:7:8$ resonant orbit.
\end{figure}

\begin{figure}
\epsscale{0.8}
\plotone{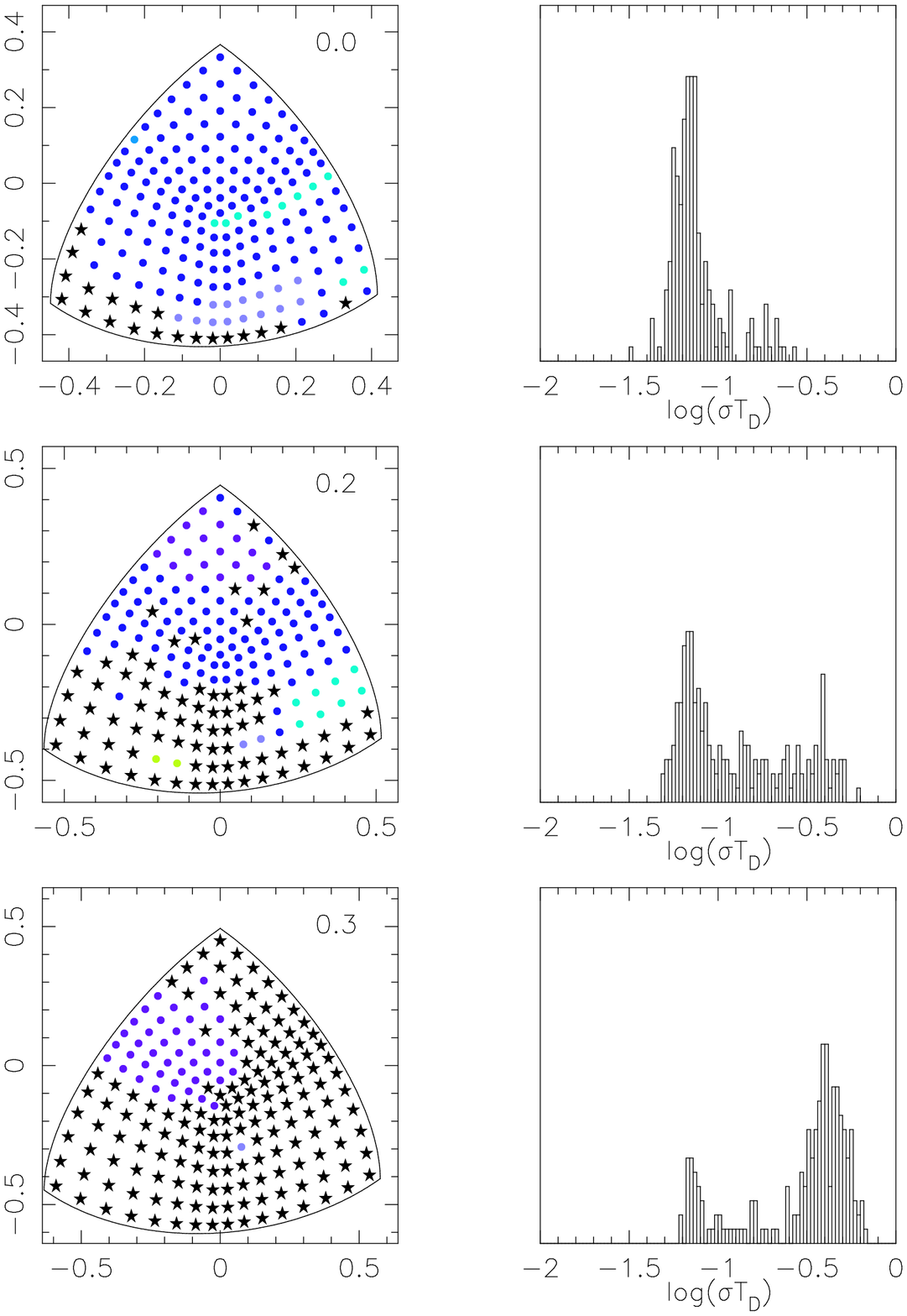} 
\figcaption[fig_stat1.ps]{\label{fig_stat1}}
Stationary start space for mass model with $\gamma=1$ (weak cusp)
and $T=c/a=0.5$. Left panels show initial positions on one octant
of the equipotential surface; $x$ axis is toward the lower left and
$z$ axis is up. Frames are labelled by $\logm$.
Circles represent starting points of regular orbits and stars
represent stochastic orbits. Colors match the colors of the 
orbit families in Figure 1. Right panels show histograms
of Liapunov exponents computed over 100 dynamical times.
\end{figure}

\begin{figure}
\epsscale{0.8}
\plotone{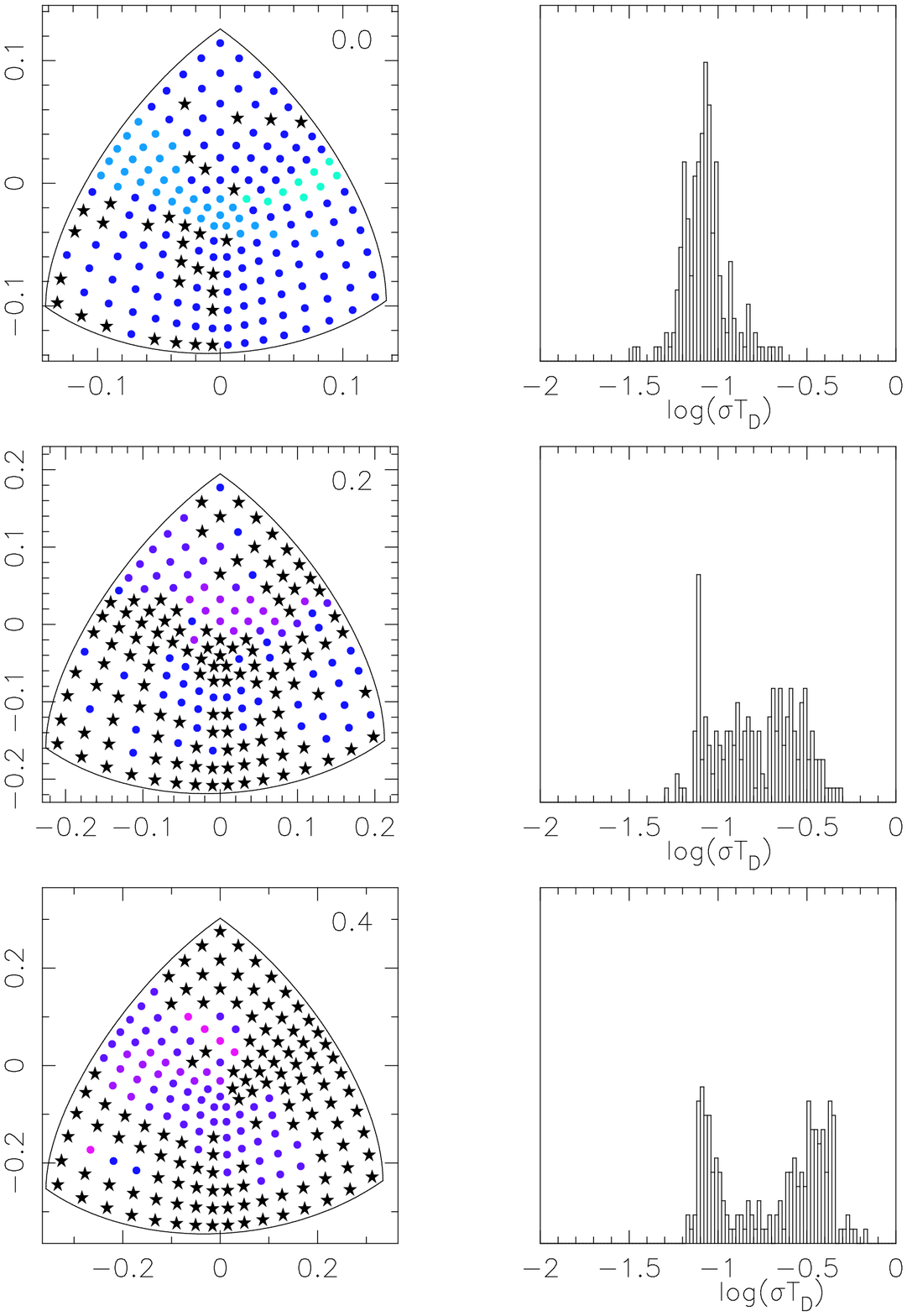}
\figcaption[fig_stat2.ps]{\label{fig_stat2}}
Like Figure 2, for $\gamma=2$ (strong cusp). As in the weak-cusp case,
pyramid orbits dominate at low energies and bananas at high energies.
\end{figure}

\begin{figure}
\epsscale{0.7}
\plotone{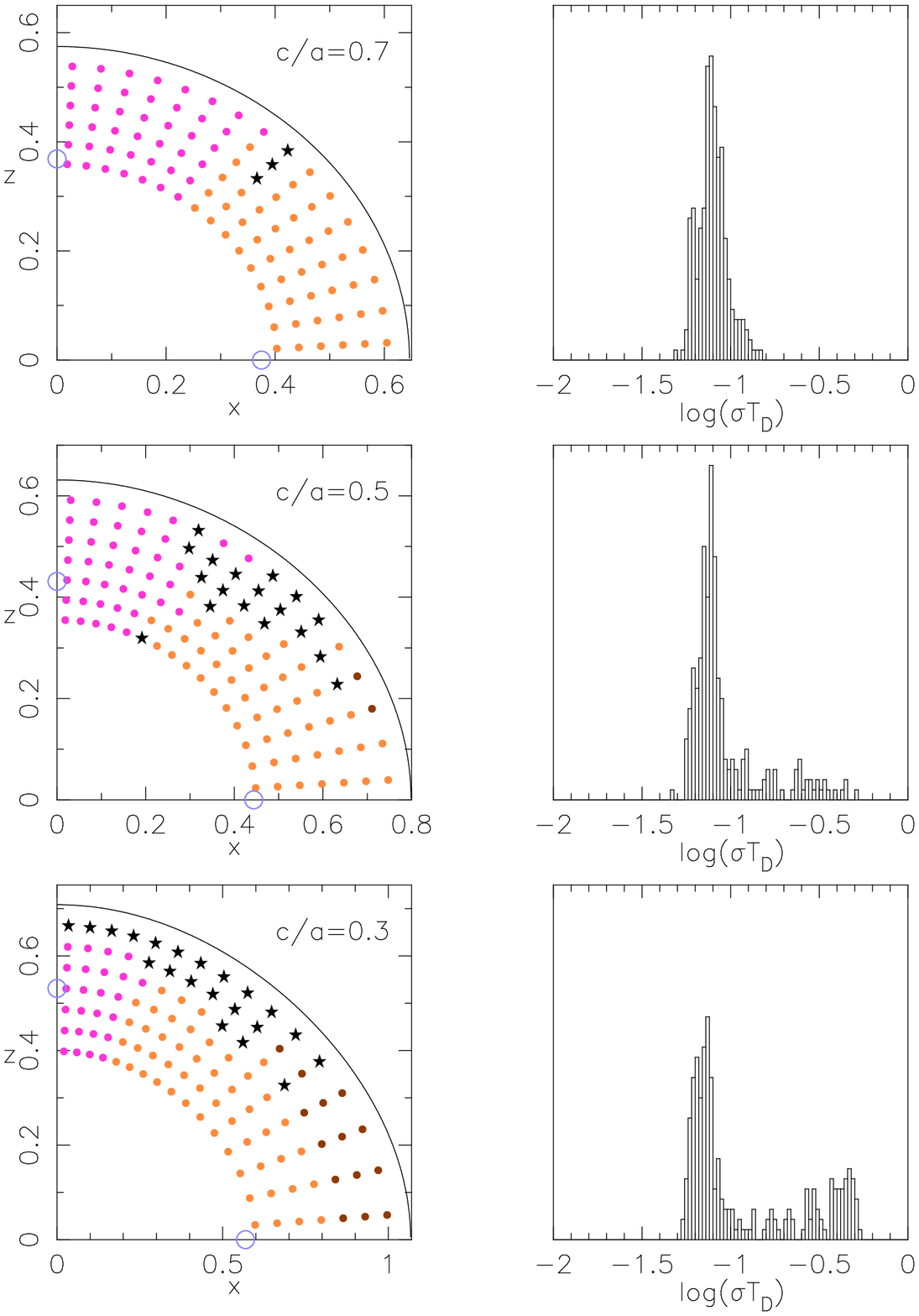}
\figcaption[fig_xz.ps]{\label{fig_xz}}
$X-Z$ start space for mass models with $\gamma=1$ (weak cusp),
$T=0.5$ and three values of $c/a$. Energy is fixed to that
of the shell with $\logm=0.2$. Left panels show initial positions in the 
($x,z$) plane. 
Circles represent starting points of regular orbits and stars
represent stochastic orbits. Colors match the colors of the 
orbit families in Figure 1. 
Open circles mark the $1:1$ closed orbits in the principal planes.
Right panels show histograms
of Liapunov exponents computed over 100 dynamical times.
As the elongation of the model is increased, stochastic and
resonant orbits (e.g. the saucers) become more prominent.
\end{figure}

Figures \ref{fig_orbs} - \ref{fig_xz} show the major families of 
orbits and their starting points in our triaxial potentials.
Stochastic orbits are present at all energies in both start spaces.
They are most prevalent in stationary start space, particularly
from starting points near the $x-$ and $y-$ axes. Motion in the vicinity
of the $z$ (short) axis tends to be stable. In $X-Z$ start space, 
stochastic orbits are mostly associated with starting points near the 
zero-velocity curve, 
or in a few cases with the transition regions between the different 
families of tube orbits.
As the energy is increased, stationary start space becomes more and more
dominated by stochastic orbits; this transition is discussed in more 
detail in the next section.

Regular motion in stationary start space is dominated by the pyramid orbits.
Sridhar \& Touma (1999) first described similar orbits, which they
called ``lenses,'' in planar, harmonic-oscillator potentials containing
central point masses. Merritt \& Valluri (1999) demonstrated the existence 
of the corresponding 3D orbits in triaxial black-hole potentials.
Pyramid orbits can be described as Keplerian ellipses with one focus 
lying near the black hole, and which precess in $x$ and $y$ due to torques 
from the background stellar potential.  They have a roughly rectangular 
base whose dimensions are fixed by the amplitudes of oscillation in $x$ 
and $y$. At low energies, pyramids with a range of shapes exist,
having bases elongated parallel to both the $x-$ and $y-$ axes. However 
their major elongation (more precisely, the elongation of a symmetrical 
pair of pyramid orbits oriented above and below the $x-y$ plane) 
is parallel to the $z$ (short) axis. This fact makes pyramid
orbits less useful than classical box orbits for reinforcing the
shape of the figure.

Close inspection of the pyramid orbits in our numerical integrations
reveals many resonant pyramid families.  Two of the most important are 
shown in Figure \ref{fig_orbs}: 
a $(3,0,-4)$ resonance between motion in $x$ and $z$, and a $(0,6,-5)$
resonance between motion in $y$ and $z$.

As the energy is increased, a $2:1$ resonance appears in the narrowest 
pyramid orbits lying near the short ($z$) axis. The opening angle of these 
``banana'' orbits increases rapidly with increasing energy, 
as the stationary point moves along the equipotential surface from the 
short to the long ($x$) axis. Many of the orbits from the banana family 
are found to be associated with a second resonance.  Two such, 
doubly-resonant familes are illustrated in Figure \ref{fig_orbs}:
the $(2:3:4)$ (banana-fish) and $(3:4:6)$ (banana-pretzel) orbits.

\begin{figure}
\epsscale{0.8}
\plotone{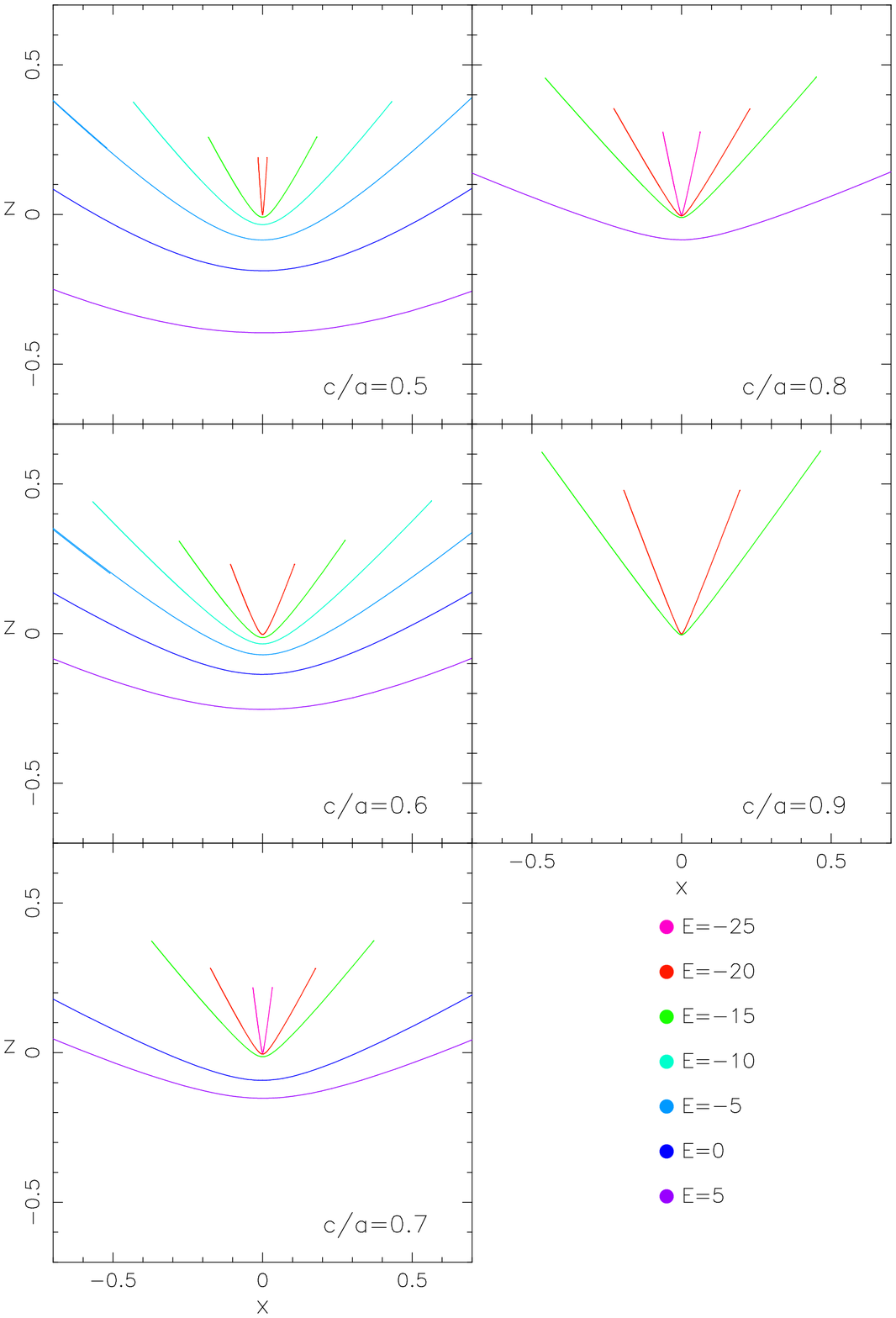}
\figcaption[fig_ban.ps]{\label{fig_ban}}
Banana orbits as a function of energy in five models,
each with $\gamma=2$ and $T=0.5$.
Unstable bananas are not shown.
\end{figure}

The variation in shape of the bananas with energy is shown in Figure 
\ref{fig_ban}. As discussed in the next section, the bananas sometimes 
persist throughout the ``zone of chaos'' and sometimes disappear, 
then reappear at high energies, with their major elongation parallel 
to the $x$-axis.  Inside of the chaotic zone, their elongation is
counter to that of the figure.

One resonant family that is apparently not associated with
the pyramids is the $(1,-2,1)$ family first discussed by Merritt 
\& Valluri (1999).
The major elongation of these orbits is parallel to the intermediate 
($y$) axis; they appear both in the weak- and strong-cusp potentials.

The orbit families in $X-Z$ start space are very similar to those in 
triaxial potentials without black holes (\cite{dez85}; \cite{sch93}; 
\cite{mef96}). Tube orbits avoid the center due to a primary, $1:1$ 
resonance in one of the principal planes and are relatively 
unaffected by the presence of the black hole.  The inner long-axis tubes 
are important only in nearly prolate potentials. The most important 
resonance among the tube orbits in our potentials is the
$2:1$ resonance in the meridional plane which generates saucer orbits
(\cite{les92}).  The saucers appear most prominantly in highly
flattened potentials.

We note an interesting feature of the motion in our models.
The dynamical roles of the long and short axes at low energies
are approximately reversed compared to their roles at large energies,
or in triaxial potentials without black holes. 
The major families of regular, boxlike orbits near the black hole -- 
the pyramids and the bananas -- are generated from Keplerian ellipses 
oriented along the short ($z$) axis,
while in triaxial potentials without central black holes, 
it is the long ($x$) axis orbit that generates the boxes and bananas. 
Similarly, stochastic 
orbits in our models derive mostly from starting points near the 
$(x-y)$ plane, while in non-singular potentials the instability strip lies
near the $(y-z)$ plane (\cite{gs81}).  
This reversal
is important because it means that most of the regular orbits near the 
black hole have the wrong elongation for supporting a triaxial 
mass distribution.

\section {Transition to Stochasticity}

The most dramatic effect of the black hole is to induce a sudden change
to stochasticity in stationary start space as the energy
is increased.  We investigated this transition as a function of $\gamma$,
$c/a$ and $T$.  Accurate orbital integrations were found to be time-consuming,
particularly at low energies and in the strong-cusp model. We therefore
used the E10K supercomputer at Rutgers University to distribute the 
computations over 64 independent processing units. At each energy in
each potential, 192 orbits were integrated starting from the equipotential
surface for a time of $100 T_D$ and the Liapunov exponents were computed.
Using all 64 processors, the elapsed time for each set of 192 orbits was 
$\sim 10$ min for $\gamma=1$ and $\sim 30$ min for $\gamma=2$.

\begin{figure}
\epsscale{0.8}
\plotone{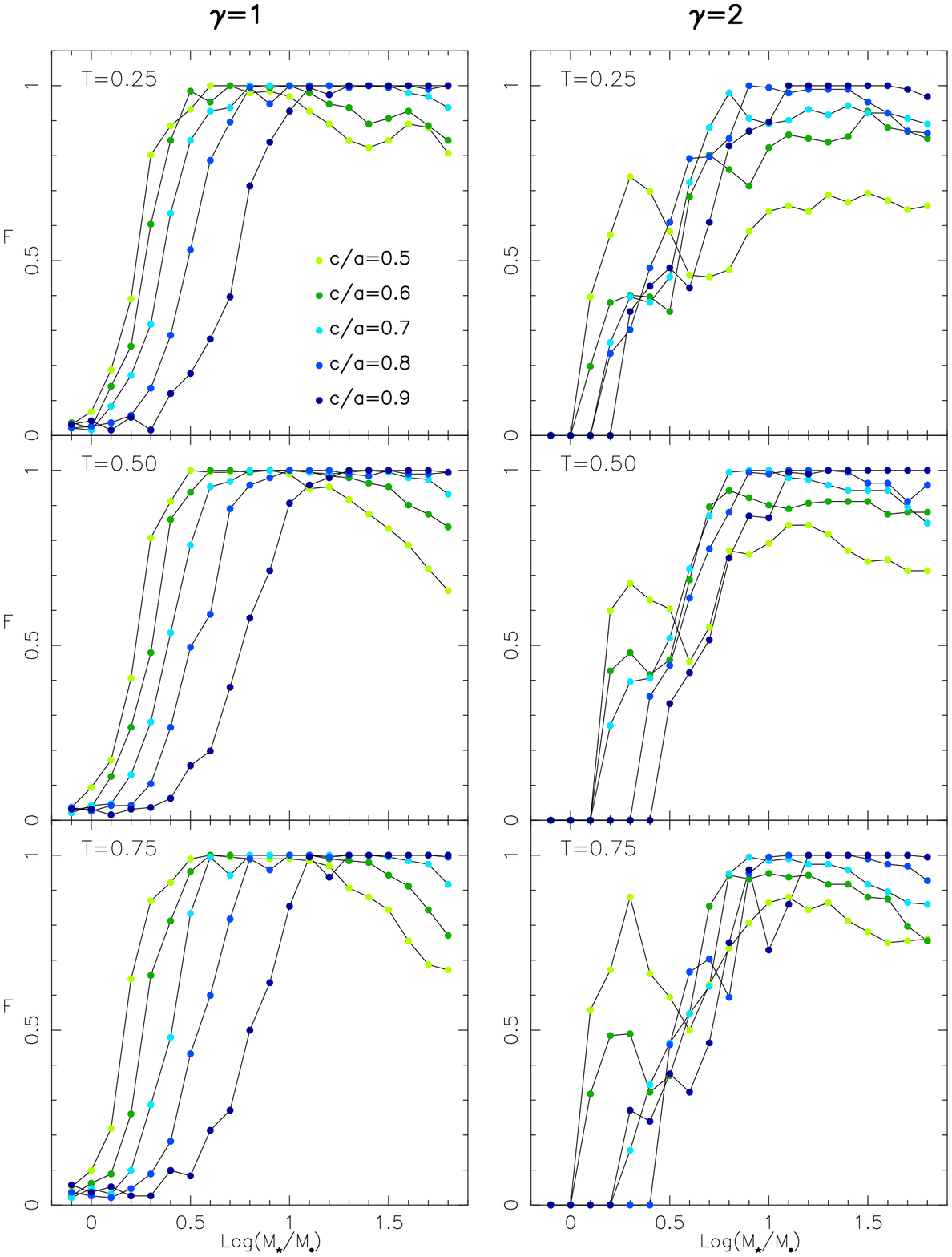}
\figcaption[fig_chaos.ps]{\label{fig_chaos}}
Fraction of chaotic orbits in stationary start space for
$\gamma=1$ (weak cusp) and $\gamma=2$ (strong cusp) 
models as a function of energy.
\end{figure}

\begin{figure}
\epsscale{0.8}
\plotone{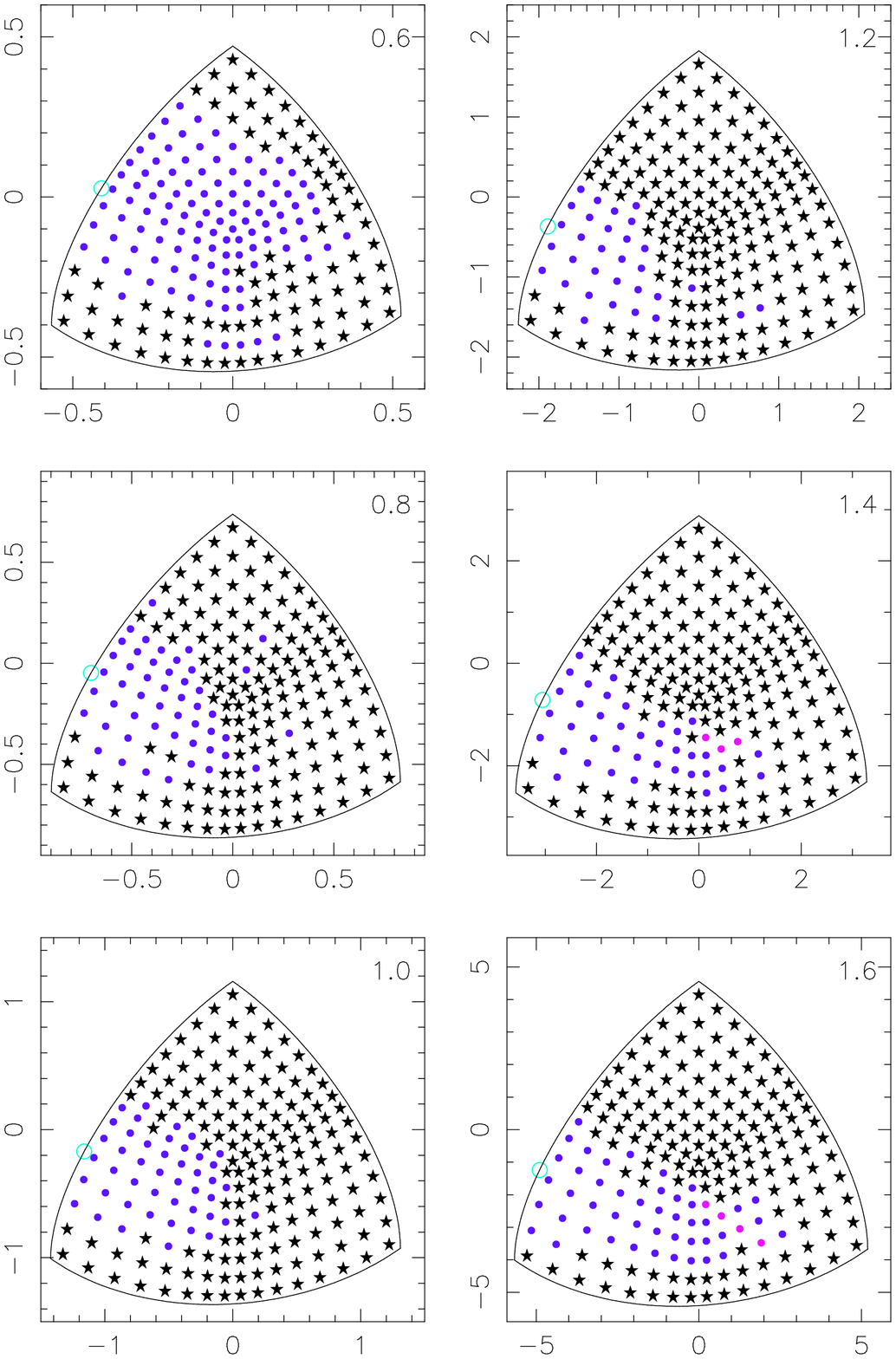}
\figcaption[fig_set.ps]{\label{fig_set}}
Stationary start space for the model with $\gamma=2, T=c/a=0.5$
as a function of energy, through the ``zone of chaos.''
The banana family of orbits persists throughout the chaotic zone.
The starting points of the resonant banana orbits are shown by the 
open circles.
\end{figure}

The results are summarized in Figure \ref{fig_chaos} and \ref{fig_set}. At each energy, 
the fraction $F$ of the 192 orbits 
that were found to be stochastic was computed and plotted. 
Stochastic orbits were identified both by their location in the histogram
of Liapunov exponents at a given energy, 
and by plots of the configuration-space trajectories.
While this fraction is not an accurate reflection of the fraction of 
phase-space associated with chaotic motion, the transition from regularity 
to chaos is so sudden that there is no need for a more accurate 
measure.

The basic character of these plots is always the same.  At low energies,
$\logm\lap 0$, the motion is almost completely regular, consisting mostly
of pyramid orbits. Starting at an energy between $\logm\approx 0$ and 
$\logm\approx 0.5$, $F$ increases suddenly to $\sim 1$ and remains
near unity over a range of energies. Finally, at high energies --
$\logm\gap 1$ for $\gamma=1$ and $\logm\gap 1.5$ for $\gamma=2$ --
$F$ begins to drop and the motion returns to a mixture of regular and 
chaotic orbits.

The existence of a ``zone of chaos'' near the centers of triaxial potentials
containing black holes was first noted by Merritt \& Valluri (1999).
Based on our more complete set of numerical experiments, we can make
the following statements about how the properties of this zone vary
with the parameters of the potential.

1. For a given triaxiality $T$, chaos sets in at lower energies in more
highly elongated models. For instance, for $T=0.5$ and $\gamma=1$, 
$F=0.8$ is reached at $\logm\approx 0.3$ for $c/a=0.5$ and $\logm\approx 0.6$
for $c/a=0.8$.

2. The transition from $F\approx 0$ to $F\approx 1$ takes place more
rapidly as a function of $\logm$ in the more elongated models.

3. The transition to chaos is interrupted by the appearance of the banana 
orbits, particularly in the more elongated potentials with $\gamma=2$.  
For instance, for $T=0.5$ and $c/a=0.5$, the chaotic
fraction first increases to $F\approx 0.7$ at $\logm=0.3$,
then decreases again to $\sim 0.4$ at $\logm=0.6$ due to the bananas
before finally increasing to $F\approx 0.8$ at $\logm\gap 1$.
The banana orbits in the most highly flattened models 
($c/a\lap 0.6$) manage to persist throughout the chaotic zone 
and keep the chaotic orbit fraction from reaching 100\%

4. The transition to chaos depends only weakly on triaxiality $T$ for
a given elongation $c/a$.

Figure \ref{fig_set} shows stationary start space as a function of
energy for a model ($\gamma=2, c/a=0.5$) in which the banana orbits persist, with an associated
family of regular orbits, throughout the zone of chaos.
In mass models where the bananas disappear, the zone of chaos
ends with the appearance of a stable resonant family
at high energies: typically the $2:3$ $x-z$ (fish) resonance for $\gamma=1$,
and the $2:1$ $x-z$ (banana) resonance for $\gamma=2$. 
At still higher energies, the orbital
populations are similar to those described by other authors
(\cite{caa98}; \cite{pal98}; \cite{vam98}; \cite{waf98}):
a complex mix of resonant box orbits, stochastic orbits, and tubes.

\section{Summary}

We have investigated the orbital motion in triaxial nuclei with 
power-law density profiles, $\rho\sim r^{-\gamma}$, 
$\gamma=(1,2)$,
and central point masses representing supermassive black holes.
The presence of the central point mass divides the phase space into
three radial regions.
At the lowest energies, the motion is essentially regular.
The major orbit families are the tubes, the pyramids, 
and a number of families associated with resonances,
most prominently the $2:1$ banana resonance.
The pyramid orbits are similar in shape to the box orbits 
of integrable triaxial potentials but have their major elongation 
parallel to the short axis, making them less useful for reconstructing
an elongated figure.
At intermediate energies, the tube orbits persist but the pyramid orbits
become increasingly chaotic.
The transition to a ``zone of chaos'' occurs rapidly in all of the
potentials investigated here, at an energy where the enclosed stellar
mass is a few times the mass of the central point. 
In the most elongated models with $\gamma=2$, the bananas can persist
throughout the zone of chaos; their axis of elongation gradually shifts
from the short axis to the long axis of the figure.
At higher energies, stable resonant boxlike orbits begin to appear in
stationary start space, generated either from closed orbits like the
fish or bananas, or from thin orbits corresponding to a 3D resonance.

Our results are limited in their applicability to the central regions
of galaxies where the stellar density profile can be approximated as
a single power law, $\rho\sim r^{-\gamma}$.  
In bright elliptical galaxies and bulges,
this is the region within $r\approx r_b$, the break radius, where the
central, shallow power law turns over to a steeper dependence at large 
radii. However we argued (Table 1) that $r_b$ is approximately
the radius within which the gravitational force from the black hole dominates
that from the stars; and the results of \S4 show that this is also
approximately the radius of transition to the ``zone of chaos'' induced
by the black hole.  Thus the onset of chaos in the phase space of
real triaxial galaxies should occur at approximately the same radius
or energy as calculated here.

Our results highlight two different ways in which central black holes
would be expected to limit the degree of triaxiality of real galactic 
nuclei.
First, the regular orbits associated with motion within the ``zone of
chaos,'' the pyramids and the tubes, are mostly poorly suited to reinforcing 
the major elongation of the figure.
Second, the black hole induces chaos in the motion of filled-center orbits
like the pyramids, causing them to occupy a region that is rounder than
that defined by the equidensity contours of the model.
We would therefore expect the degree of triaxiality to be limited inside 
the zone of chaos by the shapes of the regular orbits,
and within this region by the lack of regular orbits.
These expectations will be tested in a future study where self-consistent
triaxial models will be constructed.

\bigskip\bigskip

This work was supported by 
NSF grant AST 96-17088 and NASA grant NAG 5-6037.


\begin{deluxetable}{lcccc}
\tablewidth{0pt}
\tablenum{1}
\tablecaption{Structural Parameters for Two Elliptical Galaxies\label{T1}}
\tablehead{
\colhead{Galaxy} & 
\colhead{$\gamma$} & 
\colhead{$r_b$ (pc)} &
\colhead{$\mh (\Msolar) $} & 
\colhead{$M_*(r_b) (\Msolar) $}
}
\startdata
NGC 3379 & $1.07$\tablenotemark{(1)} & $51$\tablenotemark{(1)}  & $1.35\times 10^8$\tablenotemark{(2)} & $3.46\times 10^8$ \\
M87 & $1.26$\tablenotemark{(3)} & $315$\tablenotemark{(3)} & $3.6\times 10^9$\tablenotemark{(4)}  & $4.75\times 10^9$ 
\enddata
\tablenotetext{1}{\cite{geb96}}
\tablenotetext{2}{\cite{geb00}}
\tablenotetext{3}{\cite{vdm94}}
\tablenotetext{4}{\cite{mac97}}
\end{deluxetable}

\begin{deluxetable}{llll}
\tablewidth{0pt}
\tablenum{2}
\tablecaption{Energy Shells for Weak Cusp Potential, $T=c/a=0.5$\label{T2}}
\tablehead{
\colhead{$x_*$} &
\colhead{$E$} &
\colhead{$\log \left(\frac{M_{\star}}{M_{\bullet}}\right)$} &
\colhead{$T_D$}
}
\startdata
0.5655 & 0.3514 & ~~-0.100 & 0.8521 \\ 
0.6345 & 0.8023 & ~~0.000 & 0.9823 \\ 
0.7120 & 1.2639 & ~~0.1000 & 1.1249 \\ 
0.7988 & 1.7423 & ~~0.2000 & 1.2790 \\ 
0.8963 & 2.2437 & ~~0.3000 & 1.4436 \\ 
1.0057 & 2.7750 & ~~0.4000 & 1.6172 \\ 
1.1284 & 3.3430 & ~~0.5000 & 1.7987 \\
1.2661 & 3.9555 & ~~0.6000 & 1.9869 \\ 
1.4205 & 4.6204 & ~~0.7000 & 2.1811 \\ 
1.5939 & 5.3466 & ~~0.8000 & 2.3806 \\ 
1.7884 & 6.1438 & ~~0.9000 & 2.5855 \\ 
2.0066 & 7.0224 & ~~1.0000 & 2.7958 \\ 
2.2514 & 7.9943 & ~~1.1000 & 3.0120 \\ 
2.5261 & 9.0723 & ~~1.2000 & 3.2348 \\ 
2.8344 &10.2706 & ~~1.3000 & 3.4651 \\ 
3.1802 &11.6052 & ~~1.4000 & 3.7038 \\ 
3.5682 &13.0938 & ~~1.5000 & 3.9521 \\ 
4.0036 &14.7562 & ~~1.6000 & 4.2108 \\ 
4.4922 &16.6144 & ~~1.7000 & 4.4814 \\ 
5.0403 &18.6930 & ~~1.8000 & 4.7648 \\ 
\enddata
\end{deluxetable}

\begin{deluxetable}{llll}
\tablewidth{0pt}
\tablenum{3}
\tablecaption{Energy Shells for Strong Cusp Potential, $T=c/a=0.5$\label{T3}}
\tablehead{
\colhead{$x_*$} &
\colhead{$E$} &
\colhead{$\log \left(\frac{M_{\star}}{M_{\bullet}}\right)$} &
\colhead{$T_D$}
}
\startdata
0.1599 &-23.6463 & ~~-0.1000 &0.1268  \\ 
0.2013 &-20.8513 & ~~0.0000 &0.1721  \\
0.2534 &-18.3209 & ~~0.1000 &0.2315  \\ 
0.3191 &-16.0006 & ~~0.2000 &0.3090  \\ 
0.4017 &-13.8471 & ~~0.3000 &0.4090  \\
0.5057 &-11.8263 & ~~0.4000 &0.5371  \\
0.6366 &-9.9108  & ~~0.5000 &0.7003  \\
0.8015 &-8.0789  & ~~0.6000 &0.9070  \\
1.0090 &-6.3135  & ~~0.7000 &1.1679  \\
1.2702 &-4.6008  & ~~0.8000 &1.4961  \\
1.5991 &-2.9301  & ~~0.9000 &1.9082  \\ 
2.0132 &-1.2927  & ~~1.0000 &2.4248  \\ 
2.5344 & 0.3183  & ~~1.1000 &3.0717  \\ 
3.1907 & 1.9083  & ~~1.2000 &3.8812  \\ 
4.0168 & 3.4815  & ~~1.3000 &4.8936  \\ 
5.0569 & 5.0415  & ~~1.4000 &6.1595  \\
6.3662 & 6.5910  & ~~1.5000 &7.7420  \\
8.0146 & 8.1321  & ~~1.6000 &9.7199  \\
10.0897& 9.6666  & ~~1.7000 &12.1919 \\ 
12.7022&11.1958  & ~~1.8000 &15.2814 \\ 
\enddata
\end{deluxetable}

%
%


\begin{thebibliography}{}

\bibitem[Bender \& Saglia 1999]{bes99} Bender, R. \& Saglia, R. 1999,
	in Galaxy Dynamics,
	ASP Conf. Ser. 182, ed. D. Merritt, J. A. Sellwood \& 
	M. Valluri (ASP: Provo), 113
\bibitem[Carlson 1988]{car88} Carlson, B. C. 1988, A Table of
	Elliptic Integrals of the Third Kind. Mathematics of
	Computation, Vol. 51, No. 183, 267
\bibitem[Carpintero \& Aguilar 1998]{caa98} Carpintero, D. D. \&
	Aguilar, L. A. 1998, MNRAS, 298, 1
\bibitem[Chandrasekhar 1969]{cha69} Chandrasekhar, S. 1969,
	Ellipsoidal Figures of Equilibrium [Dover: New York]
\bibitem[Crane et al. 1993]{cra93} Crane, P. et al. 1993,
	AJ, 106, 1371
\bibitem[Dehnen 1993]{deh93} Dehnen, W. 1993, MNRAS, 265, 250
\bibitem[de Zeeuw 1985]{dez85} de Zeeuw, P. T. 1985, MNRAS, 216, 273
\bibitem[de Zeeuw \& Pfenniger 1988]{dep88} de Zeeuw, P. T. \&
	Pfenniger, D. 1988, MNRAS, 235, 949
\bibitem[Ebisuzaki, Makino \& Okumura 1991]{emo91} Ebisuzaki, T.,
	Makino, J. \& Okumura, S. K. 1991, Nature, 354, 212
\bibitem[Ferrarese et al. 1994]{fer94} Ferrarese, L., van den Bosch, 
	F. C., Ford, H. C., Jaffe, W. \& O'Connell, R. W. 1994,
	AJ, 108, 1598
\bibitem[Ford et al. 1998]{for98} Ford, H. C., Tsvetanov, Z. I., Ferrarese, L.
	\& Jaffe, W. 1998, in IAU Symp. 184, The Central Regions of the 
	Galaxy and Galaxies, ed. Y. Sofue (Dordrecht: Kluwer), 377
\bibitem[Gebhardt et al. 1996]{geb96} Gebhardt, K. et al. 1996,
	AJ, 112, 105
\bibitem[Gebhardt et al. 2000]{geb00} Gebhardt, K. et al. 2000,
	AJ, 119, 1157
\bibitem[Gerhard \& Binney 1985]{gb85} 
	Gerhard, O. E. \& Binney, J. 1985, MNRAS, 216, 467
\bibitem[Goodman \& Schwarzschild 1981]{gs81} 
	Goodman, J. \& Schwarzschild, M. 1981, ApJ, 245, 1087
\bibitem[Hairer \& Wanner 1996]{haw96} Hairer, E. \& Wanner, G.
	1996, Solving Ordinary Differential Equations. II.
	[Berlin: Springer]
\bibitem[Kormendy 1985] {k85} Kormendy, J. 1985, ApJ, 292, L9
\bibitem[Lees \& Schwarzschild 1992]{les92} Lees, J. F. \&
	Schwarzschild, M. 1992, ApJ, 384, 491
\bibitem[Macchetto et al. 1997]{mac97} Macchetto, F. et al. 1997,
	ApJ, 489, 579
\bibitem[Makino 1997]{mak97} Makino, J. 1997, ApJ, 478, 58
\bibitem[Merritt 1999]{mer99} Merritt, D. 1999, in Galaxy Dynamics,
	ASP Conf. Ser. 182, ed. D. Merritt, J. A. Sellwood \& 
	M. Valluri (ASP: Provo), 164
\bibitem[Merritt \& Fridman 1995]{mef95} Merritt, D. \&
	Fridman, T. 1995, in Fresh Views of Elliptical Galaxies,
	ASP Conf. Ser. 86, ed. A. Buzzoni, A. Renzini \& A. Serrano
	(ASP: Provo), 13
\bibitem[Merritt \& Fridman 1996]{mef96} Merritt, D. \&
	Fridman, T. 1996, ApJ, 460, 136
\bibitem[Merritt \& Valluri 1996]{mev96} Merritt, D. \&
	Valluri, M. 1996, ApJ, 471, 82 
\bibitem[Merritt \& Valluri 1999]{mev99} Merritt, D. \&
	Valluri, M. 1999, AJ, 118, 1177 
\bibitem[Papaphilippou \& Laskar 1998]{pal98} Papaphilippou, Y. \&
	Laskar, J. 1998, A\& A, 329, 451
\bibitem[Peebles 1972]{pee72} Peebles, P. J. E. 1972,
	Gen. Rel. Grav., 3, 63
\bibitem[Quillen 1999]{qui99} Quillen, A. C. 1999, 
	in Galaxy Dynamics,
	ASP Conf. Ser. 182, ed. D. Merritt, J. A. Sellwood \& 
	M. Valluri (ASP: Provo), 138
\bibitem[Quinlan \& Hernquist 1997]{quh97} Quinlan,
	G. \& Hernquist, L. 1997, New A, 2, 533
\bibitem[Quinlan, Hernquist \& Sigurdsson 1995]{qhs95} Quinlan,
	G., Hernquist, L. \& Sigurdsson, S. 1995, ApJ, 440, 554
\bibitem[Ryden 1999]{ryd99} Ryden, B. 1999, 
	in Galaxy Dynamics,
	ASP Conf. Ser. 182, ed. D. Merritt, J. A. Sellwood \& 
	M. Valluri (ASP: Provo), 142
\bibitem[Schwarzschild 1993]{sch93} Schwarzschild, M. 1993, ApJ,
	409, 563
\bibitem[Sridhar \& Touma 1999]{srt99} Sridhar, S. \& Touma, J. 1999,
	MNRAS, 303, 483
\bibitem[Sambhus \& Sridhar 2000]{sas00} Sambhus, N. \& Sridhar, S.
	2000, preprint
\bibitem[van der Marel 1994]{vdm94} van der Marel, R. 1994, MNRAS,
	270, 271
\bibitem[Valluri \& Merritt 1998]{vam98} Vallrui, M. \& Merritt, D. 
	1998, ApJ, 506, 686 
\bibitem[Wachlin \& Ferraz-Mello 1998]{waf98} Wachlin, F. C. \&
	Ferraz-Mello, S. 1998, MNRAS, 298, 22

\end{thebibliography}
\end {document}